\documentclass[10pt,twocolumn,twoside,english,conference]{IEEEtran}

\usepackage{amsthm}
\usepackage{amsfonts}
\usepackage{amssymb}
\usepackage{latexsym}
\usepackage{cite}
\usepackage[cmex10]{amsmath}
\usepackage{algorithmic}
\usepackage{url}
\usepackage{cite}
\usepackage[tight]{subfigure}

  \usepackage[pdftex]{graphicx}

\newcommand{\proba}{\mathbf{P}}

\newcommand{\rrn}{r_{r}}

\newcommand{\vdrift}{v_{d}}
\newcommand{\vdriftx}{v_{d,x}}
\newcommand{\vdrifty}{v_{d,y}}
\newcommand{\vdriftz}{v_{d,z}}

\newcommand{\ntx}{N^{\text{Tx}}}
\newcommand{\nrx}{N^{\text{Rx}}}
\newcommand{\ntxt}[1]{N^{\text{Tx}}_{#1}}
\newcommand{\nrxt}[1]{N^{\text{Rx}}_{#1}}

\newcommand{\yis}[1]{Y_{#1}}
\newcommand{\xis}[1]{X_{#1}}

\newcommand{\xih}[2]{X_{#1:#2}}

\newcommand{\histo}{\eta}

\newcommand{\rmsediff}{\Delta}

\newcommand{\phitt}[1]{P_{#1}}

\newcommand{\thr}{\xi}
\newcommand{\ts}{t_s}

\newcommand{\ith}{i^{th}}
\newcommand{\kth}{k^{th}}

\begin{document}
%
\title{Arrival Modeling and Error Analysis for Molecular Communication via Diffusion with Drift}


\author{
\IEEEauthorblockN{H.~Birkan~Yilmaz and Chan-Byoung~Chae}
\IEEEauthorblockA{School of Integrated Technology, \\
Yonsei Institute of Convergence Technology, \\
Yonsei University, Korea\\
Email: \{birkan.yilmaz, cbchae\}@yonsei.ac.kr}
\and
\IEEEauthorblockN{Burcu~Tepekule and Ali~E.~Pusane}
\IEEEauthorblockA{Department of Electrical and Electronics Engineering,\\ Bogazici University, Istanbul, Turkey\\
Email: \{burcu.tepekule, ali.pusane\}@boun.edu.tr}}


\maketitle

\begin{abstract}
\boldmath The arrival of molecules in molecular communication via diffusion (MCvD) is a counting process, exhibiting by its nature binomial distribution. Even if the binomial process describes well the arrival of molecules, when considering consecutively sent symbols, the process struggles to work with the  binomial cumulative distribution function (CDF). Therefore, in the literature, Poisson and Gaussian approximations of the binomial distribution are used. In this paper, we analyze these two approximations of the binomial model of the arrival process in MCvD with drift. Considering the distance, drift velocity, and the number of emitted molecules, we investigate the regions in which either Poisson or Gaussian model is better in terms of root mean squared error (RMSE) of the CDFs; we confirm the boundaries of the region via numerical simulations. Moreover, we derive the error probabilities for continuous communication and analyze which model approximates it more accurately. 
\end{abstract}


\IEEEpeerreviewmaketitle
\section{Introduction}
Nanonetworking is a new communication paradigm that involves various communication methods that can be used to transmit information between micro- and/or nano-scale nodes~\cite{nakano2013molecularC}. Molecular communication is a suitable method for communication at this scale, where  information is transferred not with electromagnetic or acoustic waves but molecules. In the literature, various molecular communication systems, such as molecular communication via diffusion (MCvD), calcium signaling, microtubules, pheromone signaling, and bacterium-based communication have been  proposed~\cite{nakano2013molecularC, nakano2012molecularCA,akyildiz2011nanonetworksAN}. Among these systems, one notable for beeing an effective and energy-efficient method is MCvD~\cite{kuran2010energyMF}. In MCvD, the transmitter modulates  the information onto one of the physical properties of the information carrying molecules, such as concentration, type, etc., and these molecules propagate through the environment following the physical characteristics of the medium. During the intended symbol slot, only some of these molecules arrive at the receiver (i.e., hit the receiver), and the properties of these received molecules constitute the received signal. The remaining molecules have the potential to hit during the subsequent symbol slots and cause inter-symbol interference (ISI)~\cite{yilmaz2014simulationSO}. To demodulate the received signal properly, the physical properties of the received molecules must be accurately modeled.

In the literature, the simplest and most commonly used modulation technique is the binary concentration shift keying (BCSK). In BCSK, the amplitude of the signal is considered to be the number of the received molecules in a time slot~\cite{kuran2011modulationTF}. Therefore, modeling this number and the process is a crucial point in analytical studies. Considering the reception of molecules as the success probability, the number of received molecules exhibits, naturally, a binomial process. When multiple emissions are considered, due to the ISI caused by the diffusion channel, previous transmissions must also be taken into account for the determination of the current symbol. Considering the effects of the previously transmitted symbols requires a summation of the binomial random variables, which is analytically hard to work with. Therefore, in the literature, two approximations of the binomial distribution are used, namely the Poisson and Gaussian approximations~\cite{kilinc2013receiverDF, mahfuz2014strengthBO, noel2014improvingRP, kuran2010energyMF, yilmaz2014arrivalMF}.

In this paper, Poisson and Gaussian approximation models are compared for a given set of system parameters, such as transmitter-receiver distance, drift velocity, and the number of molecules emitted from the transmitter. As performance metrics, this study utilizes bit error rates and root mean squared error (RMSE) of cumulative distribution functions (CDF). RMSE and bit error rate plots are divided into two regions, representing which approximation is better than the other. Analytical results are verified by Monte Carlo simulations for a 3-dimensional (3-D) environment.

The remainder of the paper is organized as follows. In Section II, the MCvD system model is introduced, and BCSK is briefly explained. The arrival model of molecules and analytical calculation of error probabilities are given in Section III. In Section IV, numerical results of the performance evaluations are presented, and Section V concludes the paper.

\section{System Model}

\subsection{Molecular Communication via Diffusion}
\begin{figure}[t]
\centering
\includegraphics[width=1.0\columnwidth,keepaspectratio]
{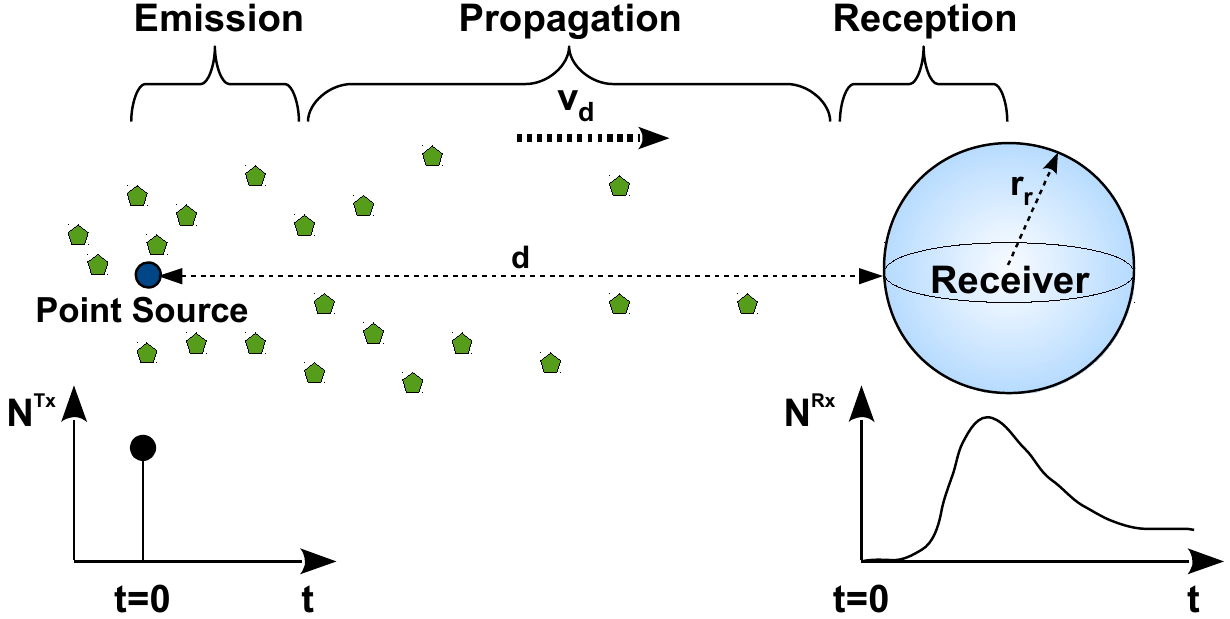}
\caption{System model of MCvD with drift.}
\label{fig_mcvd_system_model}
\end{figure}
The MCvD system with drift velocity $\vdrift$ is depicted in Fig.~\ref{fig_mcvd_system_model}. Four essential entities are the source, information molecules, receiver, and the aqueous medium. The spherical receiver and the point source are located in an aqueous medium that has a flow and the messenger molecules are the information particles in molecular scale. The distance between the receiver and the point source is denoted by $d$. The received and the sent signals are also depicted in Fig.~\ref{fig_mcvd_system_model}. The point source releases $\ntx$ molecules at time $t=0$ and, at the receiver side, time versus the number of received molecules ($\nrx$) is depicted as the received signal.

Three main processes of the system are emission, propagation, and reception. Each particle released from the point source has two components in its movement dynamics: one originates from the Brownian motion while the other one arises from the flow. The propagation model with drift is, thus, given~as
\begin{align}
\begin{split}
(x_{t}, y_{t}, z_{t}) &= (x_{t-\Delta t}, y_{t-\Delta t}, z_{t-\Delta t}) + (\Delta x, \Delta y, \Delta z)\\
\Delta x & \sim \mathcal{N}(0, 2D\Delta t) +  \vdriftx \,\Delta t\\
\Delta y & \sim \mathcal{N}(0, 2D\Delta t) +  \vdrifty \,\Delta t\\
\Delta z & \sim \mathcal{N}(0, 2D\Delta t) +  \vdriftz \,\Delta t
\end{split}
\label{eqn_propagation_model}
\end{align}
where $\Delta x_{t}$, $\Delta y_{t}$, $\Delta z_{t}$, $D$, $\vdriftx$, $\vdrifty$, $\vdriftz$, and $\mathcal{N}(\mu, \sigma^2)$ are the particle's position at each dimension at time $t$, the diffusion coefficient, the drift velocities at each dimension, and the normal distribution with mean $\mu$ and the variance $\sigma^2$, respectively.

During its trip, a molecule can hit the spherical receiver with radius $\rrn$ and, in this case, the hitting molecule is absorbed by the spherical receiver. Therefore, a molecule can contribute to the signal just once. The propagation formulations of the system change compared to the system without drift, due to the different propagation dynamics. Therefore, the closed form solution for the expected fraction of molecules hitting to the receiver in a 3-D environment that was derived  in~\cite{yilmaz2014_3dChannelCF} cannot be directly used for the current system with drift.

\subsection{Modulation and Demodulation}
In this communication system, information is sent using a sequence of symbols that are spread over sequential time slots ($\ts$) with one symbol in each slot. The symbol sent by the transmitter is called the ``intended symbol" and the $\ith$ intended symbol is denoted by $\xis{i}$. The demodulated symbol at the receiver is called the ``received symbol". 

The task of the molecule emission process is to modulate the information on some of the physical properties of the molecule. Concentration shift keying (CSK) is one of the proposed modulation techniques in the nanonetworking domain~\cite{kuran2011modulationTF}. Binary CSK (BCSK) is the most commonly used and the simplest modulation technique for MCvD~\cite{kuran2011modulationTF}. In this paper, we consider BCSK modulation, i.e., $\ntx$ molecules are emitted for a bit-1 and no emission is done for a bit-0.

Demodulation takes place within the reception process and BCSK symbols are demodulated by thresholding the number of received molecules in a given symbol slot. The demodulation function, $\delta(.)$, receives  $\nrx$ as an input and outputs the demodulated bits according to
\begin{align}
\yis{i} = \delta(\nrxt{i}) = 
\left\{ 
	\begin{array}{ll}
	0 & \mbox{if } \nrxt{i} \leq \thr \\
	1 & \mbox{if } \nrxt{i} > \thr 
	\end{array}
\right.
\end{align}
where $\nrxt{i}$ and $\yis{i}$ denote the number of received molecules and the demodulated symbol in the $\ith$ symbol duration, respectively,  while $\thr$ denotes the threshold value for demodulation.

\section{Arrival Modeling \& Error Probabilities}
The arrival of molecules in an MCvD system obeys, by its nature, the binomial distribution by considering the hitting probability as the success probability. If we consider a single emission of $\ntx$ molecules at $t=0$,  $\nrxt{1}$ exhibits a binomial random variable given as
\begin{equation}
\nrxt{1} \sim \mathcal{B}(\ntx, \phitt{1}),
\end{equation}
where $\phitt{1}$ denotes the expected fraction of molecules absorbed by the receiver node during the first symbol duration, while $\mathcal{B}(n,\,p)$ denotes the binomial distribution with $n$ trials and a success probability $p$. 

Hitting probabilities describe the mean channel response, which implies that the choice of the symbol duration has a great significance in the  determination of the channel response. It is desirable to have the first hitting probability $\phitt{1}$ to be the largest in magnitude compared to $\phitt{i}$ for $i>1$ to reduce the inter-symbol interference.

For the general case with multiple emissions, the number of received molecules in a time slot is affected by the current and the previous emissions. Therefore, we obtain 
\begin{equation}
\nrxt{i} \sim \displaystyle\sum\limits_{k=1}^{i} \mathcal{B}(\ntxt{k}, \phitt{i-k+1}),
\end{equation}
where $\ntxt{k}$ denotes the number of emitted molecules in the $\kth$ symbol duration.

It is, however, hard to work with the binomial CDF when considering consecutively sent symbols as we need a summation of random variables. Therefore, in the literature, two approximations of the binomial distribution are used, namely, Poisson and Gaussian approximations~\cite{kuran2011modulationTF, kuran2010energyMF, kim2013novelMT, noel2014improvingRP}.

\subsection{Modeling $\nrxt{i}$}
\begin{figure}[t]
\centering
\includegraphics[width=1\columnwidth,keepaspectratio]
{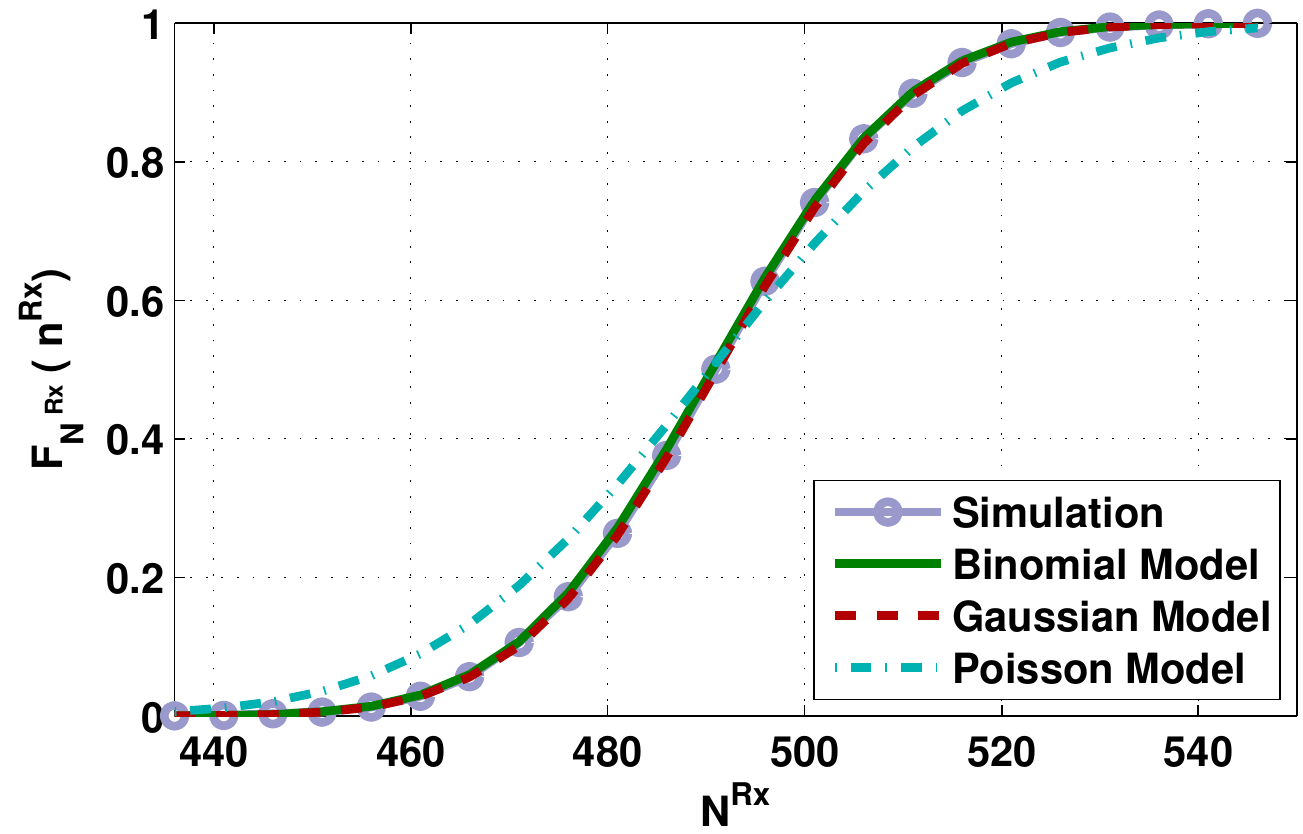}
\caption{Simulation and the model CDFs for single emission ($\ntx = 1000$, $d=5\,\mu m$, $\rrn = 10 \,\mu m$, $D=79.4 \,\mu^2/s$, $\vdrift=10\,\mu m/s$, $\ts=0.4\,s$).}
\label{fig_cdf_for_nrxi}
\end{figure}

The arrival of molecules is a binomial process in its nature. The intractability of binomial random variables, however, suggests using approximations; this model is mostly approximated by the Poisson model given as 
\begin{equation}
\nrxt{i} \sim \mathcal{P}(\sum\limits_{k=1}^{i} \ntxt{k}\,\phitt{i\!-\!k+1})
\end{equation}
or by the Gaussian model, given as
\begin{align}
\nrxt{i} \sim \mathcal{N}\left(\sum\limits_{k=1}^{i} \ntxt{k}\phitt{i\!-\!k\!+\!1},\,  \sum\limits_{k=1}^{i} \ntxt{k}\phitt{i\!-\!k\!+\!1}(1\!-\!\phitt{i\!-\!k\!+\!1}) \right)
\end{align}
where $\mathcal{P}(\lambda)$ stands for the Poisson distribution with mean $\lambda$. 

We investigate which model is better for representing the process with given parameters. For this purpose, we developed a custom simulator for MCvD with drift in MATLAB. It simulates the propagation model explained in (\ref{eqn_propagation_model}) and removes the arriving particles (i.e.,  particles that hit the receiver) from the environment. As expected, for a given set of parameters, at each run, $\nrxt{i}$ has a different value and the simulator outputs these values for CDF estimation. $\nrxt{i}$ values are utilized to estimate the CDF $F_{\nrxt{i}}(x)$ using  
\begin{equation}
\label{eqn_cdf_estimation}
F_{\nrxt{i}}(x) = \proba(\nrxt{i} \leq x)
\end{equation}
where $\proba(.)$ stands for the event probability. 

We compare the CDFs of models with the simulation output in Fig.~\ref{fig_cdf_for_nrxi}. We simulated a single emission of 1,000 molecules 90,000 times and evaluated the estimated CDF. The binomial and Gaussian models are relatively acceptable models while the Poisson model deviates considerably from the estimated CDF. 

To measure the difference between CDFs of $\nrx$ for a single emission, we evaluate the RMSE between a model and the estimated CDFs as
\begin{align}
\label{eqn_rmse_formulation}
\text{RMSE} &= \sum\limits_{x=0}^{\ntx} 
\sqrt{\frac{\left(F_{\nrxt{i}}^{\text{sim}}(x) - F_{\nrxt{i}}^{\text{model}}(x)\right)^2}{\ntx+1}},
\end{align}
where $F_{\nrxt{i}}^{\text{sim}}(x)$ and $F_{\nrxt{i}}^{\text{model}}(x)$ are the CDFs obtained from the simulations and the analytical model, respectively.

\subsection{Error Probabilities}
When there is a continuous transmission of bits, $\nrxt{i}$ is affected by the current and the previous emissions which are determined by the transmitted bit values. In this paper, for the error analysis, we consider BCSK modulation with no emission for bit-0 and the demodulation is carried out via simple thresholding. 

Considering the sequence of previously transmitted bits, $\xih{1}{i}$, we obtain the error probabilities at the $\ith$ symbol slot for a given $X_i$  as
\begin{align}
\label{eqn_error_prob_given_xihZero}
\begin{split}
\proba_{e|X_i=0,\xih{1}{i\!-\!1}} &= \proba(Y_i\!=\!1 \,| \, X_i\!=\!0, \, \xih{1}{i\!-\!1})\\
 			&= \proba(\nrxt{i} > \thr\,| \, X_i\!=\!0, \, \xih{1}{i\!-\!1})
\end{split}
\end{align}
\begin{align}
\label{eqn_error_prob_given_xihOne}
\begin{split}
\proba_{e|X_i=1,\xih{1}{i\!-\!1}} &= \proba(Y_i\!=\!0 \,| \, X_i\!=\!1, \, \xih{1}{i\!-\!1})\\
 			&= \proba(\nrxt{i} \leq \thr\,| \, X_i\!=\!1, \, \xih{1}{i\!-\!1})
\end{split}
\end{align}
where $\proba_{e|X_i,\xih{1}{i\!-\!1}}$ denotes the conditional probability of error for the $\ith$ bit with $X_i$ and $\xih{1}{i\!-\!1}$ given. 

We utilize \eqref{eqn_error_prob_given_xihZero} and \eqref{eqn_error_prob_given_xihOne} to evaluate the average error probability over all possible bit sequences of $\xih{1}{i}$ and obtain
\begin{align}
\label{eqn_average_err_prob_last}
\begin{split}
\proba_{e} 
 &= \sum\limits_{\xih{1}{i\!-\!1}} \proba_{e|X_i=0,\xih{1}{i\!-\!1}} \, 
 \proba(X_i=0)\proba(\xih{1}{i\!-\!1})\\
 &+ \sum\limits_{\xih{1}{i\!-\!1}} \proba_{e|X_i=1,\xih{1}{i\!-\!1}} \, 
 \proba( X_i=1)\proba(\xih{1}{i\!-\!1}).
\end{split}
\end{align}
We assume that the current symbol is affected by at most $\histo$ previous symbols and utilize \eqref{eqn_average_err_prob_last} to calculate the error probability depending on $\histo$ and the system parameters. Most prior work showed that, with an appropriate symbol duration, the current symbol is mostly affected by one previous symbol \cite{kuran2010energyMF, kuran2011modulationTF, kim2014symbolIO, arjmandi2013diffusionBN}. Therefore, the value of $\histo$ does not need to be large.


\section{Results and Discussion}
\begin{figure*}[t]
\begin{center}
\subfigure[$\vdrift = 0 \, \mu m/s$]
{\includegraphics[width=0.49\textwidth,keepaspectratio]
{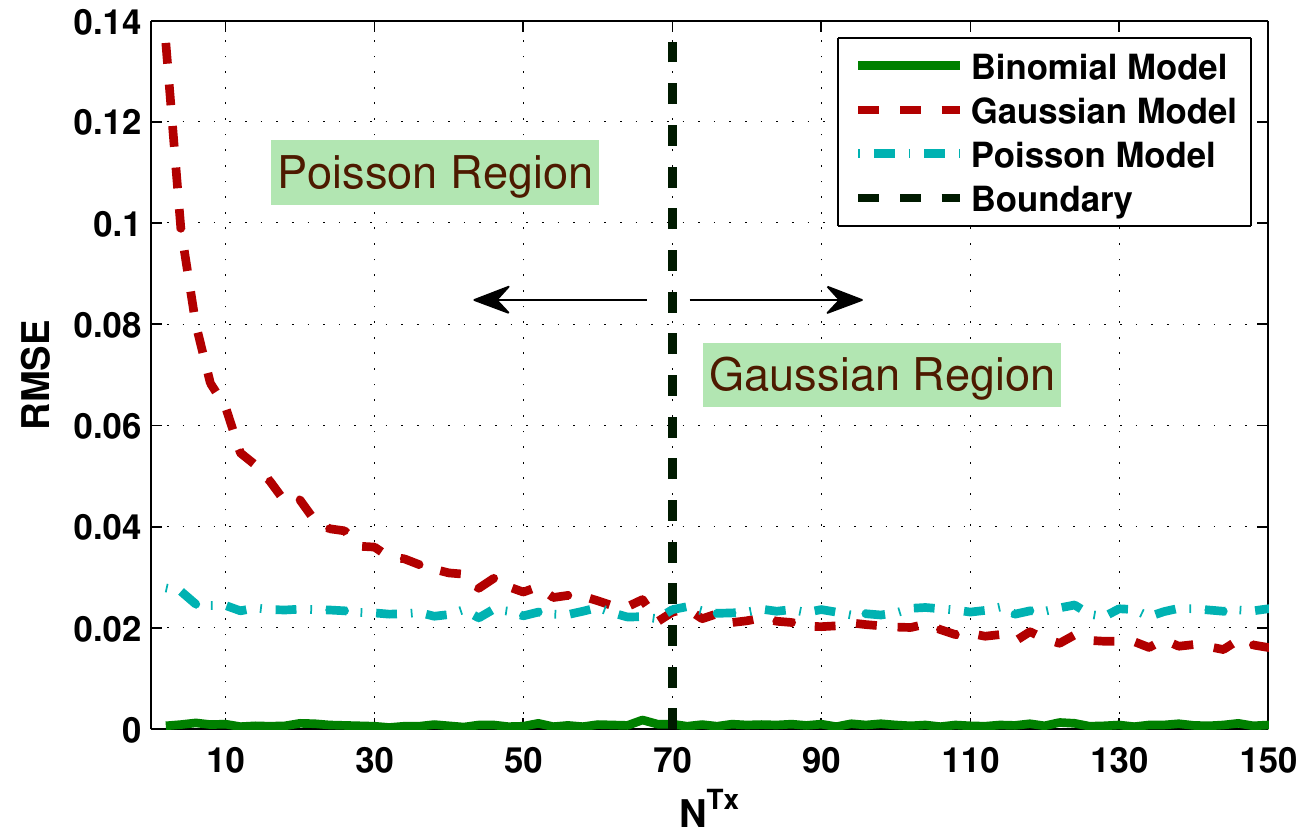}
\label{subfig_ntx_vs_rmse_woutDrift} } 
\subfigure[$\vdrift = 10 \, \mu m/s$]
{\includegraphics[width=0.49\textwidth,keepaspectratio]
{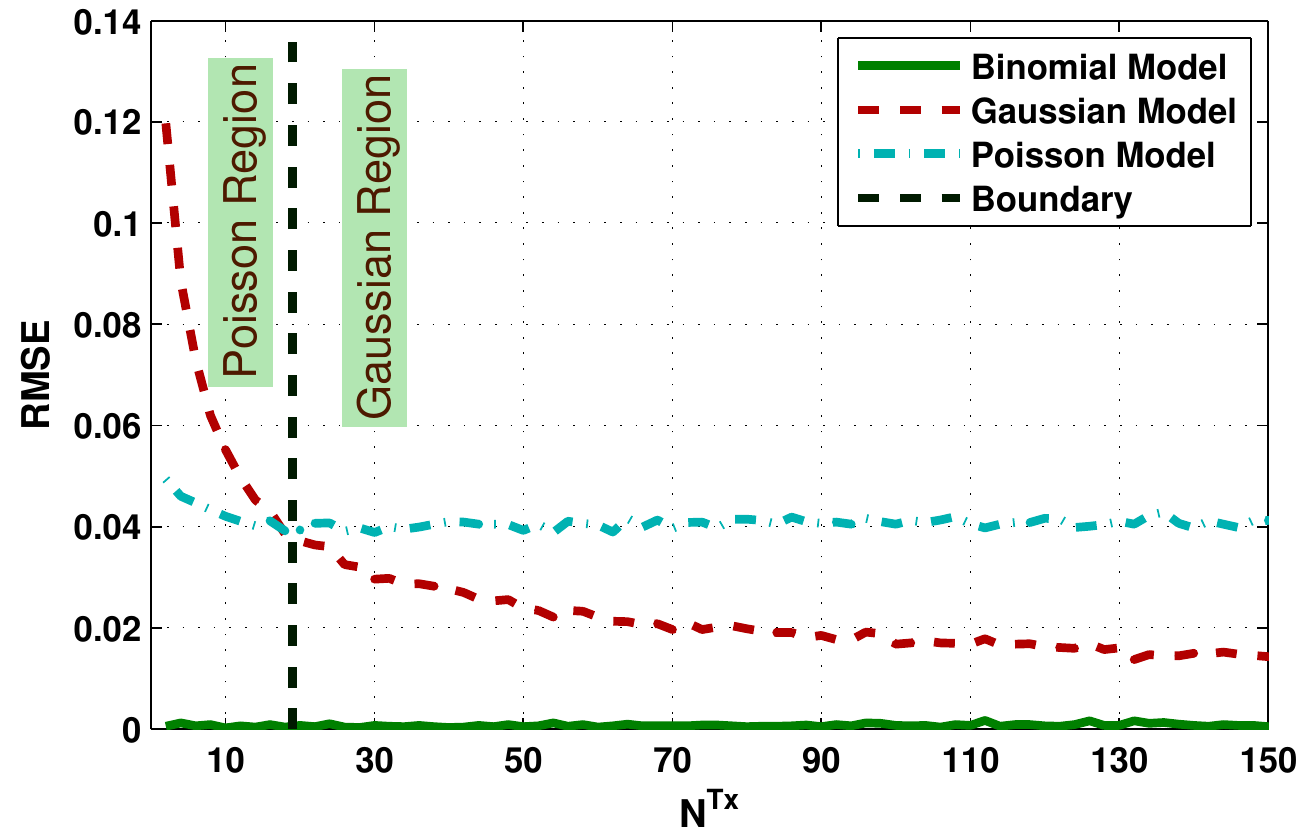}
\label{subfig_ntx_vs_rmse_withDrift} }
\end{center}
\caption{Effect of $\ntx$ on RMSE of CDFs ($d=6\,\mu m$, $\ts=0.5\,s$).}
\label{fig_ntx_vs_rmse}
\end{figure*}

\subsection{Performance Metrics and Parameters}
We use the RMSE defined in \eqref{eqn_rmse_formulation} as the main performance metric and analyze the distance, $\ntx$, and the velocity parameters to determine the best choice for the approximation under given parameters. After investigating the RMSE of CDFs, we then focus on the  probability of errors. Each approximation yields to its own $\proba_e$ due to different CDFs and we compare the results with simulation data.

In Table~\ref{tbl_system_parameters}, we present the system parameters and their values or ranges that we used for the experiments presented throughout the paper. 
\begin{table}[h]
\begin{center}
\caption{Range of parameters used in the analysis}
\renewcommand{\arraystretch}{1.2}
\label{tbl_system_parameters}
\begin{tabular}{p{5cm} l}
\hline
\bfseries{Parameter} 							& \bfseries{Value} \\ 
\hline 
 Diffusion coefficient ($D$) 		& $79.4 ~ (\mu m)^2 / s $ \\ 
 Radius of the receiver ($\rrn$)   	& $10 ~\mu m$                \\
 Distance ($d$)					& $\{1 \sim 10\} ~\mu m$\\
 $r_0$                			& $\rrn + \{1 \sim 10\} ~\mu m$\\ 
 Drift velocity ($\vdrift$)			& $\{0 \sim 10\}~\mu m/s$\\
 $\ntx$ 						& $\{2 \sim 200, \, 1000\}$ molecules\\ 
 Detection threshold ($\thr$) 		& $\{0 \sim \ntx\}$ \\
 Symbol duration ($\ts$)			& $\{0.4, \, 0.5\}~s$\\
 $\histo$ 							& 5 \\
 \hline
\end{tabular} 
\end{center}
\end{table}
\renewcommand{\arraystretch}{1}
\subsection{RMSE of CDFs}
Fig.~\ref{fig_ntx_vs_rmse} shows the effect of $\ntx$ on the RMSE of CDFs. The binomial model has the smallest RMSE as expected and the Poisson
model is nearly stable while the RMSE of the Gaussian model improves as
$\ntx$ increases in both cases with and without drift. Increasing $\ntx$ results in a higher mean number of arriving molecules; hence, the deviation due to the problematic part of the Gaussian model (i.e., negative values) diminishes. There are two regions in both Figs.~\ref{subfig_ntx_vs_rmse_woutDrift} and \ref{subfig_ntx_vs_rmse_withDrift}, namely Poisson and Gaussian regions and we name them according to having smaller RMSE values. Introducing drift with a velocity of $\vdrift=10\, \mu m/s$ squeezes the Poisson region (i.e., Gaussian region starts from more smaller $\ntx$ value). 

\begin{figure}[t]
\begin{center}
\includegraphics[width=0.98\columnwidth,keepaspectratio]
{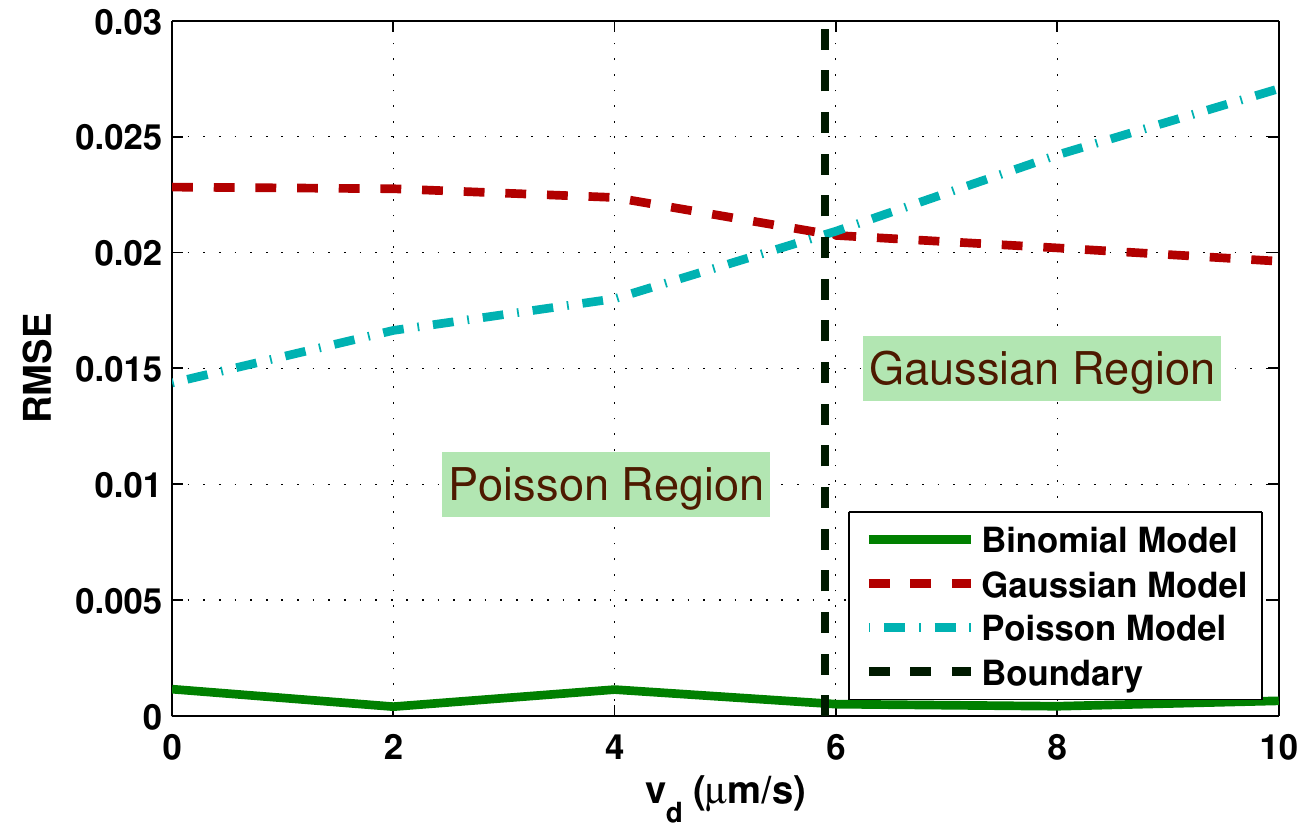}
\end{center}
\caption{Effect of $\vdrift$ on RMSE of CDFs ($\ntx = 100$, $d=8\,\mu m$, $\ts=0.5\,s$).}
\label{fig_velocity_vs_rmse}
\end{figure}

We analyze the effect of drift velocity on the RMSE of CDFs in Fig.~\ref{fig_velocity_vs_rmse}. Again, as expected, the binomial model has the smallest RMSE. Increasing drift velocity improves the accuracy of the Gaussian model slightly while deteriorating the Poisson model performance significantly. A drift velocity of approximately $\vdrift=6 \, \mu m/s$ is the turning point and after this point the Gaussian model describes the original process better.

\subsection{RMSE Heat Map}
\begin{figure}[t]
\begin{center}
\includegraphics[width=0.98\columnwidth,keepaspectratio]
{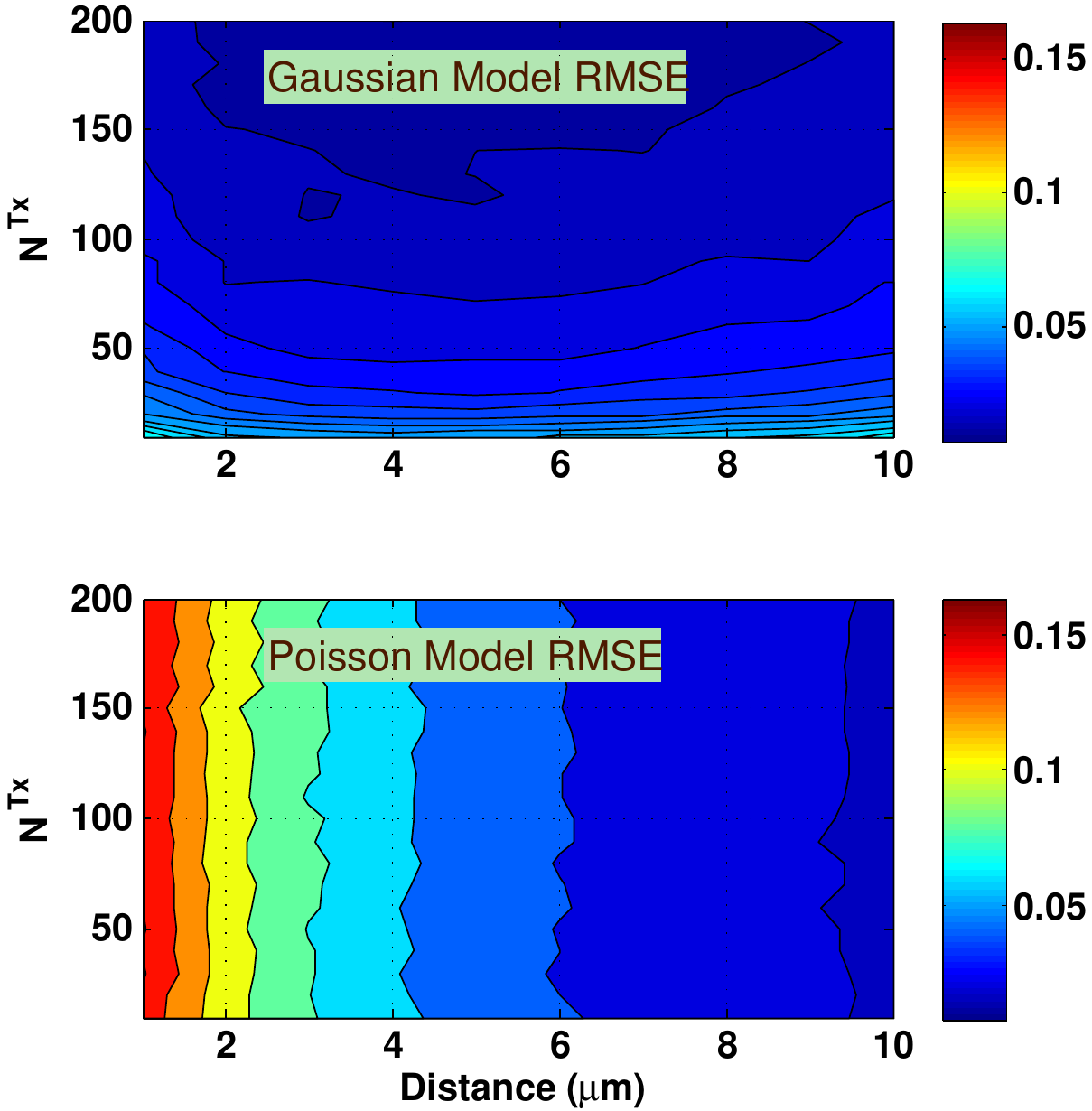}
\end{center}
\caption{RMSE heat map of CDFs (${\vdrift=10\,\mu m/s}$, $\ts=0.5\,s$).}
\label{fig_heatMapsOrig}
\end{figure}

In Fig.~\ref{fig_heatMapsOrig}, we analyze the RMSE heat map with respect to the distance and $\ntx$. For the Poisson model, the main factor is observed as the distance, which is related to the hitting probability. The Poisson distribution is better for modeling rare events~\cite{vonmises1964mathematicalTP}; hence, the significant parameter becomes the distance, since it determines the hitting probability. Therefore, for the Poisson model, we see a stable behavior for distances and better RMSE for higher distance values, since the hitting event becomes rare. On the other hand, the Gaussian approximation of binomial distribution gets better with higher $\ntx$ values and is not good for a success probability close to 0 or 1. Therefore, the RMSE increases at the higher distances and decreases for higher $\ntx$. 
\begin{figure*}[t]
\begin{center}
\subfigure[$\vdrift = 0 \, \mu m/s$]
{\includegraphics[width=0.49\textwidth,keepaspectratio]
{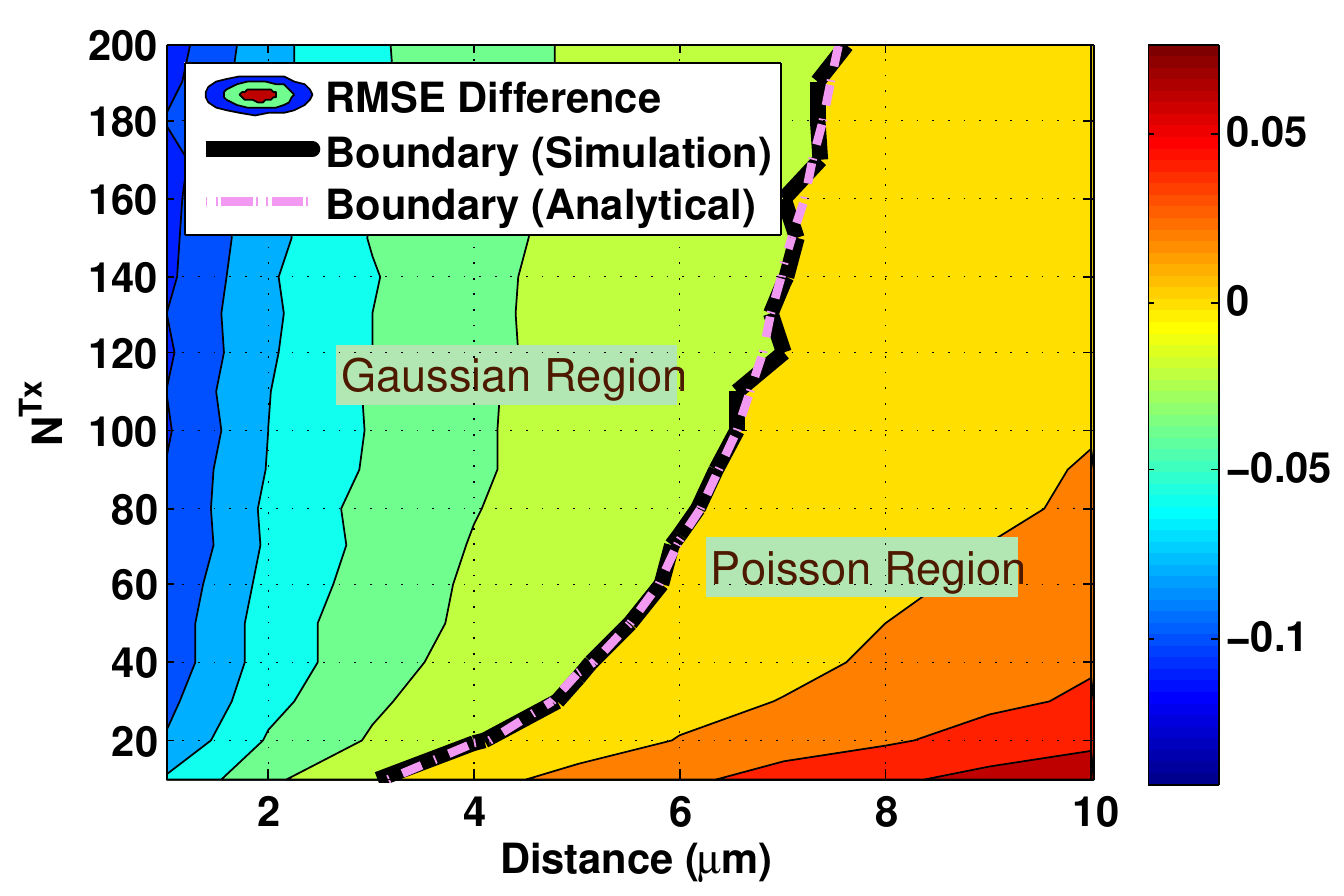}
\label{subfig_heatMaps_woutDrift} } 
\subfigure[$\vdrift = 10 \, \mu m/s$]
{\includegraphics[width=0.49\textwidth,keepaspectratio]
{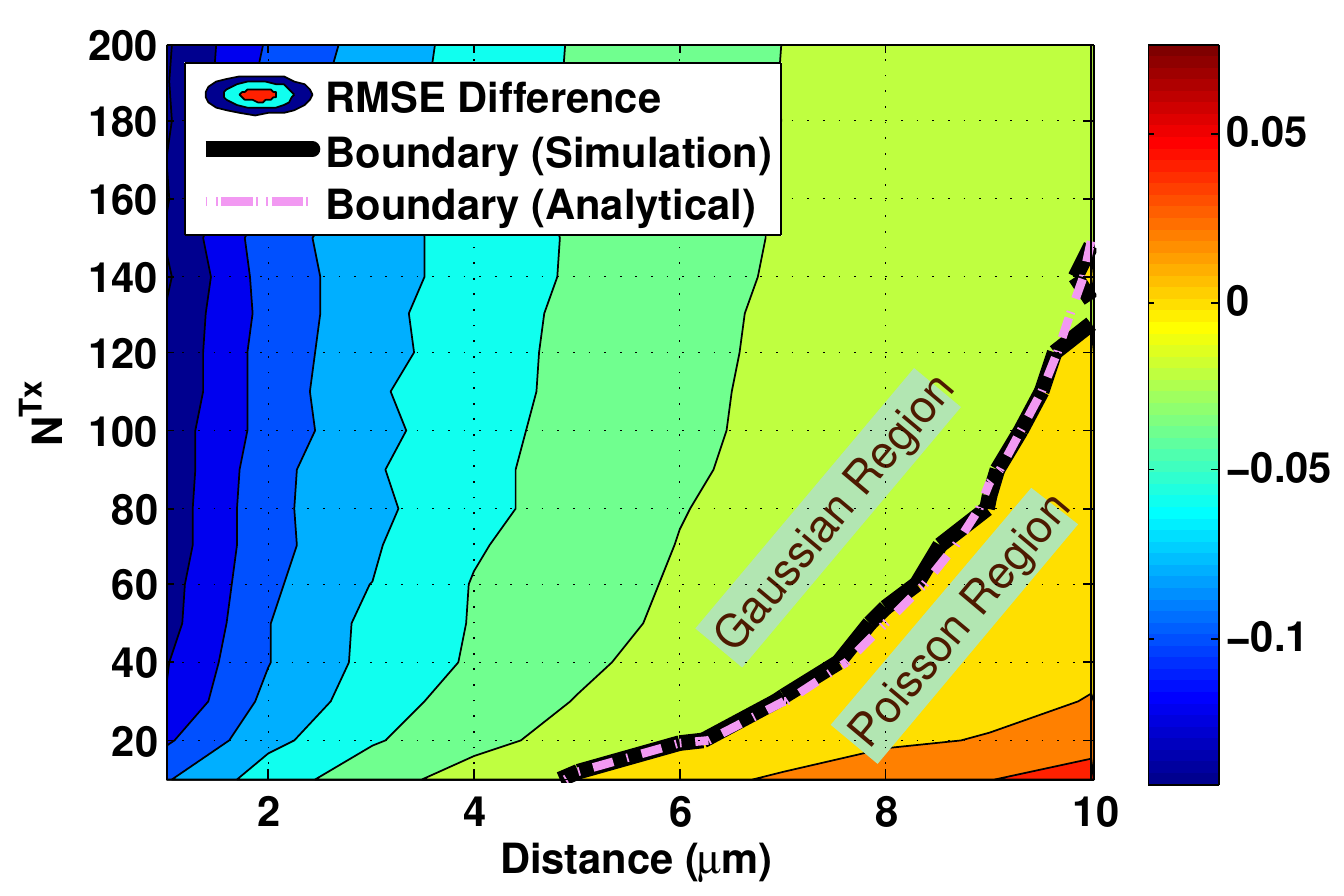}
\label{subfig_heatMaps_withDrift} }
\end{center}
\caption{RMSE difference of Gaussian and Poisson models ($\ts=0.5\,s$).}
\label{fig_heatMaps}
\end{figure*}

In Fig.~\ref{fig_heatMaps}, the heat map of RMSE difference of the Gaussian and Poisson models is depicted under both no drift and a drift velocity of $\vdrift=10\, \mu m/s$. The thicker contour corresponds to level zero and the right side of the zero level is better for the Poisson model (i.e., the RMSE of Poisson model's CDF is less than that of the Gaussian model's CDF). Introducing a drift squeezes the Poisson region and the Gaussian model starts to explain the original process better for higher distances. 

The dashed line corresponds to the analytical solution for an estimate of RMSE difference, $\rmsediff$, and it is given as
\begin{equation}
\begin{aligned}
\rmsediff &= \sqrt{\sum\limits_{x_i\in [0,\ntx]} \frac{\left[I_{1\!-\!p}(\ntx\!-\!x_i, 1\!+\!x_i)-Q\left(\frac{\ntx\,p\!-\!x_i}{\sqrt{\ntx\,p\,(1\!-\!p)}}\right)\right]^2}{\ntx+1} } \\
                         &- \sqrt{\sum\limits_{x_i\in [0,\ntx]} \frac{\left[I_{1\!-p}(\ntx\!-x_i, 1+x_i)-\frac{\Gamma(x_i+1,\ntx\,p)}{x_i!}\right]^2}{\ntx+1} } 
\label{eqn:diff_rmse}
\end{aligned}
\end{equation}
where $p=\phitt{1}$ and the functions $I_{1-p}(.,.)$ and $\Gamma(.,.)$ stand for regularized incomplete beta function and incomplete gamma function, respectively. Since the binomial model explains the arrival process perfectly, we use this fact to estimate the RMSE difference of the Gaussian and Poisson models. Equation \eqref{eqn:diff_rmse} first evaluates the RMSE difference between the Gaussian and binomial model CDFs, then between the Poisson and binomial model CDFs, and finally it calculates the difference of them.

\subsection{Error Probability Analysis}
We analyze how the deviations of CDFs impact upon $\proba_{e}$. For $\proba_{e}$ analysis, we select a point from Fig.~\ref{subfig_heatMaps_withDrift} in the Gaussian region to quantify the superiority to the Poisson model. We chose $N_1=100$ and $d=4\, \mu m$ for the analysis and considered the continuous transmission case for simulations and evaluated  $\proba_{e}$ from \eqref{eqn_average_err_prob_last} to compare the Gaussian and Poisson models. 

It is obvious that, if the detection threshold value is decreased, it is always possible to achieve a better detection performance of a bit-1. Reducing the threshold, however, may lead to incorrect demodulation of a bit-0 due to the ISI.
\begin{figure}[t]
\begin{center}
\includegraphics[width=0.98\columnwidth,keepaspectratio]
{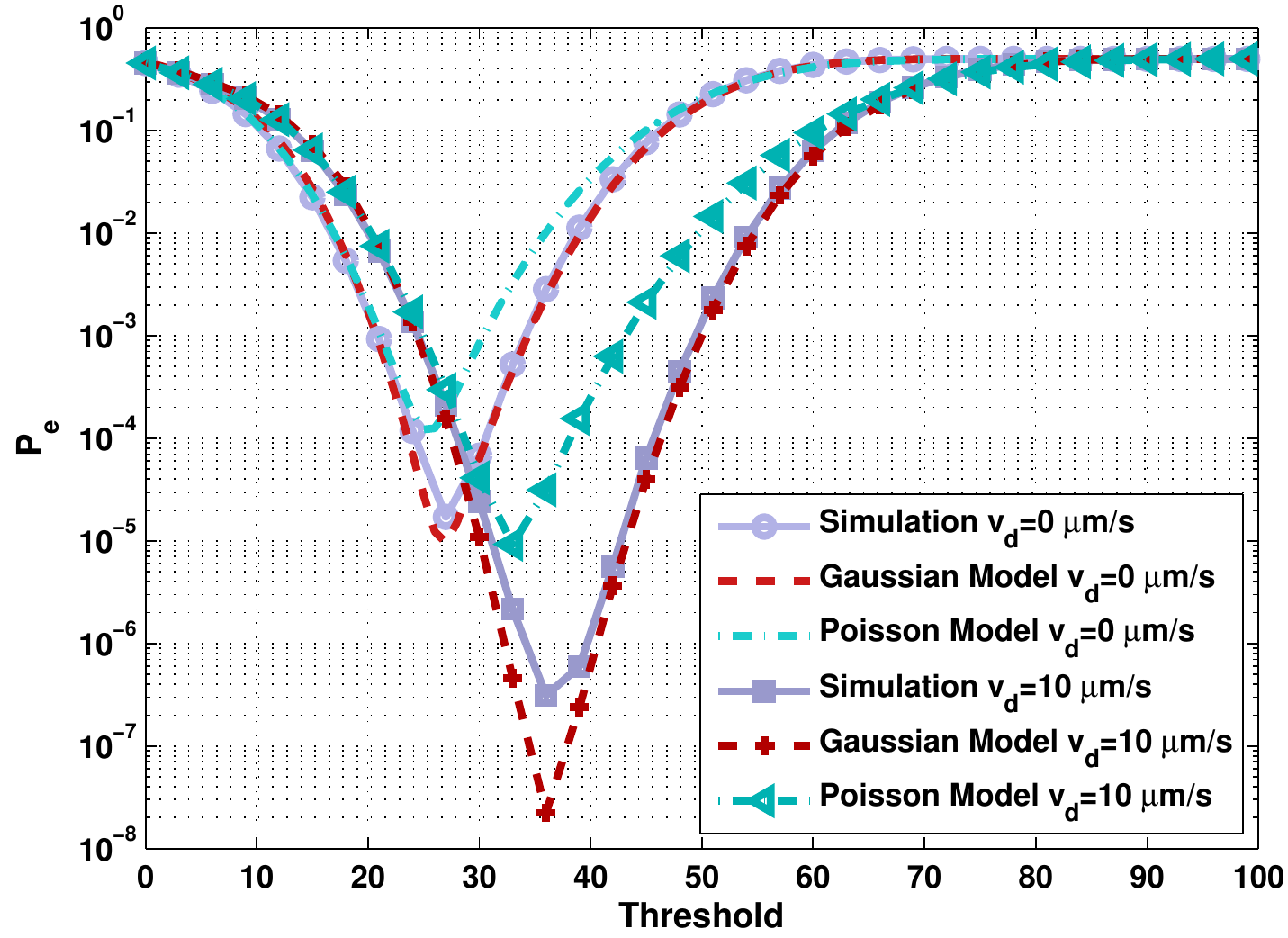}
\end{center}
\caption{Threshold versus $\proba_{e}$ ($d=4\, \mu m$, $N_0=0$ and $N_1=100$ molecules, $\ts=0.5\,s$).}
\label{fig_prob_e}
\end{figure}

In Fig.~\ref{fig_prob_e}, we analyze the error probability with and without drift. The Gaussian model defines the process better than Poisson model for the chosen parameters. Introducing drift to the system shifts the optimum threshold to higher values due to the increase in $\nrx$. In both cases, with and without drift, the Poisson model approximates  $\proba_{e}$ higher than the original value, while the Gaussian model approximates it in the opposite way with a small deviation from the actual value. Moreover, optimum threshold values agree when the process is estimated with utilizing the Gaussian approximation.

\section{Conclusion}
In this paper, we analyzed the accuracy of the arrival models for an MCvD system with drift. We determined which model approximates the original process better for the given system parameters such as drift velocity, distance, and the number of emitted molecules. In each analysis, we determined the regions of Poisson and Gaussian models in which they describe the original process better. We analytically evaluated the boundary for these regions and confirmed the results via 3-dimensional simulations. We also investigated the accuracy of the error probability $\proba_{e}$ for both Gaussian and Poisson models. Future work will consider the effects of decomposition of molecules and sparse data transmission.

\section*{Acknowledgment}
This work was in part funded by the MSIP (Ministry of Science, ICT \& Future Planning), Korea, under the ``IT Consilience Creative Program" (NIPA-2014-H0201-14-1002) supervised by the NIPA (National IT Industry Promotion Agency) and by the Basic Science Research Program (2014R1A1A1002186) funded by the MSIP, Korea, through the National Research Foundation of Korea.

\bibliographystyle{IEEEtran}
\bibliography{iccRefs}

\end{document}